\documentclass[prb,aps,twocolumn]{revtex4}
\usepackage{bm}
\usepackage{amsmath,amssymb}
\usepackage{color,multirow}
\usepackage{graphicx}
\begin{document}
\title{Theory of time reversal topological superconductivity in double
Rashba wires
-- symmetries of Cooper pair and Andreev bound states}

\author{Hiromi Ebisu$^{1}$, Bo Lu$^{1}$, Jelena Klinovaja$^{2}$, Yukio Tanaka$^{1}$}
\affiliation{$^1$~Department of Applied Physics, Nagoya University, Nagoya 464-8603, Japan\\
$^2$~Department of Physics, University of Basel, Klingelbergstrasse 82, CH-4056 Basel, Switzerland\\
}
\date{\today}
\begin{abstract}
We study the system of double Rashba wires brought into the proximity to an $s$-wave superconductor. The time reversal invariant topological superconductivity is realized if the interwire pairing corresponding to crossed Andreev reflection dominates over the standard intrawire pairing.
We derive the topological criterion and show that the system hosts zero energy Andreev bound states such as a Kramers pair of Majorana fermions. We classify symmetry of the Cooper pairs focusing on the four degrees of freedom, $i.e.$, frequency, spin, spatial parity inside wires, and
spatial parity between wires. The magnitude of the odd-frequency pairing is strongly enhanced in the topological state.
We also explore properties of junctions occurring in such double wire systems. If one section of the junction is in the topological state and the other is in the trivial state, the energy dispersion of Andreev bound states is proportional to
$\sim\pm\sin\varphi$, where $\varphi$ denotes the macroscopic phase difference between two sections. This behavior can be intuitively explained by the couplings of a Kramers pair of Majorana fermions and spin-singlet $s$-wave Cooper pair and can also be understood by analyzing an effective continuum model of the $s+p$/$s$-wave superconductor hybrid system.
\end{abstract}
\pacs{pacs}
\maketitle
\thispagestyle{empty}

\section{Introduction}

The concept of topology\cite{Schnyder08,Kitaev09} and topological effects has attracted a lot of attention over past decades. For example, the appearance of the zero energy surface Andreev bound state (SABS) in unconventional superconductors like $p$-wave \cite{ABS,ABSb}
or $d$-wave superconductors \cite{Hu,TK95,kashiwaya00}
has been understood in terms of the topological invariants defined for the
bulk Hamiltonian. \cite{STYY11,tanaka12,Kobayashi2014,Kobayashi2015}
Also the possibility to generate an effective topological $p$-wave superconductivity in the systems coupled to conventional $s$-wave superconductors due to internal spin structure\cite{Fujimoto09,lutchyn10,oreg10,alicea10,JK2012,JK2012_2,alicea12,JK2012_3,Beenakker13,JK2013,Yazdani2013,JK2014,Ebisu2} opened the field for experiments.\cite{Mourik,Deng,Rokhinson,Das,yazdaniex,Pawlak} In one dimensional systems, zero energy SABSs are Majorana fermions (MFs), which are of great importance for  topological quantum computing. \cite{Nayak,alicea11}
In this context, it is useful to shed light on MF physics from different angles. One aspect not covered in literature on MFs is the symmetry of the Cooper pairs in the topological regime.
Generally, if we consider such degree of freedom as time, space as well as spin, there are four possible symmetries of Cooper pairs: even-frequency spin-singlet even-parity (ESE), even-frequency spin-triplet odd-parity (ETO), odd-frequency spin-triplet even-parity (OTE) \cite{Berezinskii}
and odd-frequency spin-singlet odd-parity (OSO). \cite{Balatsky92,Coleman,Fuseya}
It is known that odd-frequency pairing ubiquitously exists in
inhomogeneous superconductors, \cite{odd3,tanaka12}
and it is hugely enhanced at the boundaries if zero energy SABSs are present.
\cite{odd1,odd3,odd3b,tanaka12,Eschrig2007}
The connection between MFs and odd-frequency pairing have also been clarified before in several systems. \cite{Asano2013,Wakatsuki,Ebisu,Ebisu2} It was shown  that MFs inevitably accompany odd-frequency spin-triplet pairing in the D class topological superconductors with broken time-reversal symmetry.

Alternatively, there are also time reversal invariant topological superconductors belonging to the topological DIII class\cite{Schnyder08,Kitaev09} and occuring in various condensed matter systems. \cite{Fu08,Fu10,sasaki11,hsieh12,yamakage12,JK2015,JK2015_2}
If the time reversal symmetry is not broken, two MFs come in Kramers pairs and protected from splitting.\cite{Fu09,Nakosai1,Law,Nakosai2,FuStern,Fb,JK2015_3,FanZhang,Ortiz,JK2014_2,QiJosephson,Qi09,Qi2010,Interpair}
In this work we focus on the system consisting of two quantum wires with Rashba type spin-orbit interaction (SOI) brought into proximity to an $s$-wave superconductor as introduced in Ref. [\onlinecite{Interpair}].
The interwire superconducting pairing induced by crossed Andreev processes is larger than the intrawire pairing due to strong electron-electron interactions. The crossed Andreev reflection, when two electrons forming initially the Cooper pair get separated into different channels, has attracted a special attention due to its potential use for creating entanglement\cite{JK2001,entanglement1,entanglement2} and  has been implemented in superconductor/normal metal/superconductor junctions\cite{SNStheory1,SNSexperiment1,SNSexperiment2} and double quantum dots superconductor hybrid systems.
\cite{SNStheory1,JK2001,JK2002,JK2002_2,Recher2003,SNSexperiment1,SNSexperiment2,
entanglement1,DQDexperiment2,DQDexperiment1,Ksato2010,DQDtheory1,DQDexperiment3,DQDexperiment4,
JK2012_4,entanglement2}
\par


Although the topological superconductivity and MFs have been already predicted for this model, the relation between Kramers MFs and odd-frequency pairing still remains largely unexplored for systems in the topological DIII class. For example, it is natural to expect that the spin structure of odd-frequency pairing in the presence of Kramers pairs is very different from one for the topological D class with a single MF. In addition, working with two quantum wires, we have to include one additional spatial degree of freedom such as the wire index.\cite{Blackschaffer} We will call the superconducting pairing to be wire-odd (wire-even) if the pair amplitude picks up minus (plus) sign if one exchanges two wires.
As a result, one can expect much richer structure of pairing amplitudes depending on four degrees of freedom, {\it i.e.}, frequency, spin, and the spatial parity inside the wire as well as between the two wires. Besides symmetries of Cooper pairs, the energy of Andreev bound states (ABSs) in
Josephson junctions has not been addressed in this system so far.
All this together contributes to the advancement of the
understanding of topological superconductor in DIII class and also of the
physics of pairing symmetry.
\par

The paper is organized  as follows: In Sec. \ref{sec2}, we introduce a tight-binding model of the double quantum wire (DQW) model with Rashba spin-orbit interaction and proximity induced superconductivity where both interwire and intrawire pairing potentials are taken into account. The topological criterion for the system  is analyzed beyond the linearization approximation applied in Ref. [\onlinecite{Interpair}]. We confirm that the interwire pairing potential must be larger than that of intrawire one and that the inversion symmetry should be broken. In Sec. \ref{sec3}, we study symmetries of Cooper pair of the DQW model and find various types of pair amplitudes due to the translational and inversion symmetry breaking. We show that, in topological regime, the odd-frequency pairing is strongly enhanced at the ends of the system.  We prove that the $\uparrow\downarrow$ component of the spin-triplet pairing is absent both for even and odd-frequency pairings due to the time reversal symmetry.
In Sec. \ref{sec4}, we focus on ABSs in the DQW/normal metal/DQW and
DQW/normal metal/spin-singlet $s$-wave superconductor Josephson junction system. Especially for the latter case, we see an anomalous energy dispersion  of ABSs
proportional to $\sim\pm\sin\varphi$ with  $\varphi$ being the superconducting phase difference between two sections. We provide a qualitative argument to explain this relation by considering
tunneling Hamiltonian between Kramers MFs and spin-singlet $s$-wave pairing.
We also reproduce this dispersion relation in the framework of the effective model which can be interpreted as an $s+p$/$s$-wave superconducting junction. In Sec. \ref{sec5}, we summarize our results.

\section{Model construction}

\label{sec2} 
\begin{figure}[h]
\begin{center}
\includegraphics[width=80mm]{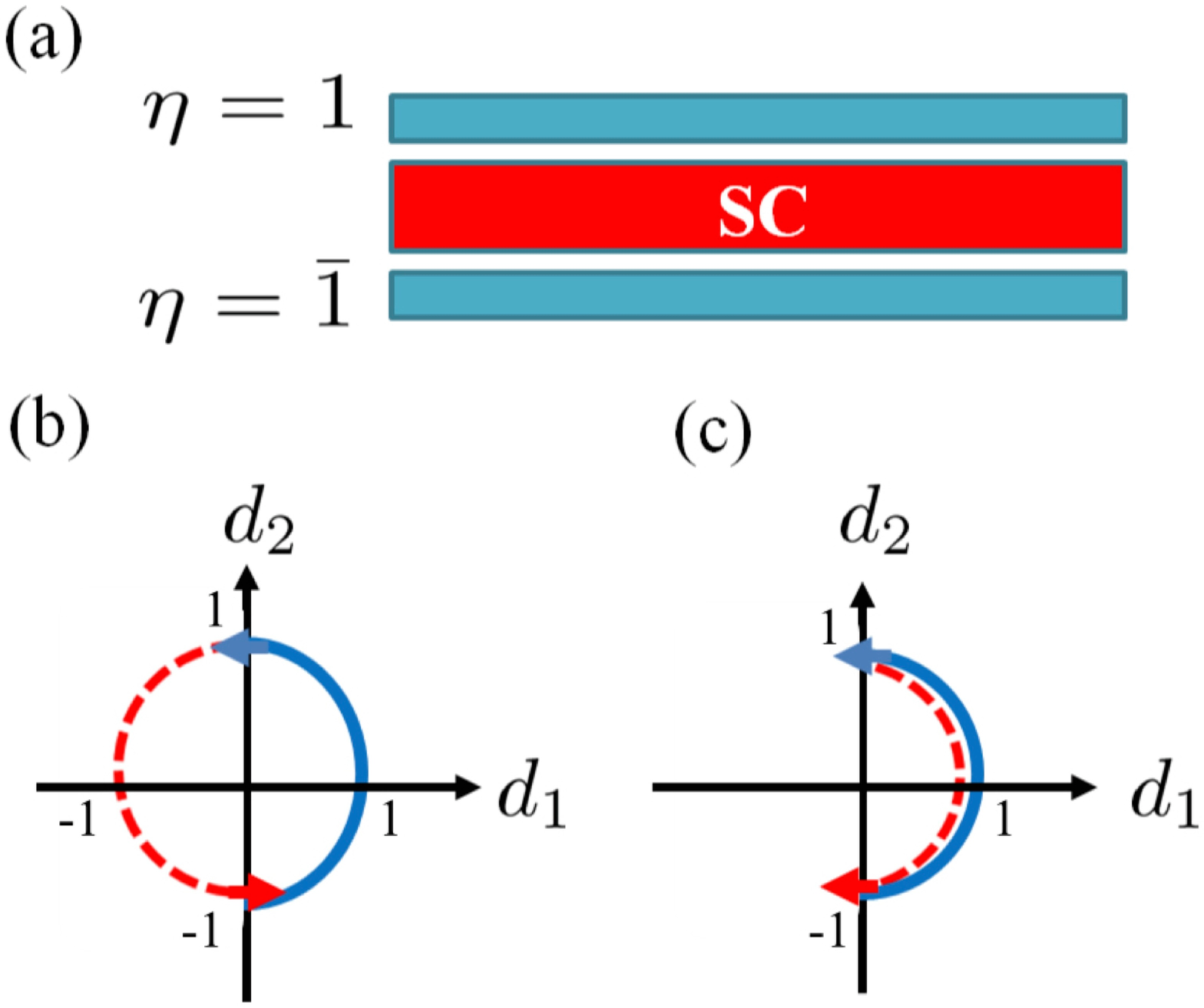}
\end{center}
\caption{(a) A schematic picture of two Rashba QWs proximity coupled with
superconductor. (b) The trajectory, spanned by the vector $(d_1,d_2)$ in the case $\text{sgn}[\protect\lambda(k_{F,1})]=-\text{sgn}[\protect\lambda(k_{F,2})]$, wraps around the origin leading to the topologically non-trivial state.
(c) The trajectory, spanned by the vector  $(d_1,d_2)$ in the case $\text{sgn}[\protect\lambda(k_{F,1})]=\text{sgn}[\protect\lambda(k_{F,2})]$, does not go around the origin which concludes that the system is topologically trivial.
}
\label{fig1}
\end{figure}

We consider the setup consisting of two quantum wires with
Rashba SOI proximity coupled to an $s$-wave superconductor (Fig. \ref{fig1}%
). The superconductivity in the DQW system could be induced in two different
ways as described in Ref. [\onlinecite{Interpair}]. The Cooper pair could
tunnel as a whole into one of two wires resulting in the intrawire
superconducting pairing. Alternatively, the Cooper pair could split such
that electrons tunnel into different wires resulting into the interwire
superconducting pairing also called the cross-Andreev superconducting
pairing. We will work in the framework of tight-binding model with the
Hamiltonian given by 
\begin{eqnarray}
H&=&-t\sum_{\left\langle
i,j\right\rangle\sigma\eta}c_{i\sigma\eta}^{\dagger}c_{j\sigma\eta}-\sum_{i%
\sigma\eta}\mu_{\eta}c_{i\sigma\eta}^{\dagger}c_{i\sigma\eta}  \nonumber \\
&+&i\sum_{\left\langle i,j\right\rangle\eta}\alpha_{\eta}\Bigl(%
c_{i\uparrow\eta}^{\dagger}c_{j\downarrow\eta}-c_{i\downarrow\eta}^{%
\dagger}c_{j\uparrow\eta}\Bigr)  \nonumber \\
&+&\sum_{i\eta}\Delta_{\eta}\Bigl(c_{i\uparrow\eta}^{\dagger}c_{i\downarrow%
\eta}^{\dagger}+\text{H.c.}\Bigr)  \nonumber \\
&+&\sum_{i}\Delta_c\Bigl(c_{i\uparrow1}^{\dagger}c_{i\downarrow\bar{1}%
}^{\dagger} +c_{i\uparrow\bar{1}}^{\dagger}c_{i\downarrow1}^{\dagger}+\text{%
H.c.}\Bigr)  \label{model}
\end{eqnarray}
Here, we introduce index $i$ ($\eta=1, \bar{1}$) to label lattice sites (QWs). We define
$c_{i\sigma\eta}^{\dagger}(c_{i\sigma\eta})$ as the creation (annihilation)
operator acting on the electron at the site $i$ of the $\eta$ QW  with the spin $\sigma$ ($\sigma=\uparrow(\equiv 1), \downarrow(\equiv \bar 1)$). The first term represents hopping with amplitude $t$ between two adjacent sites $%
\left\langle i,j\right\rangle$. The second term describes chemical potential $%
\mu_{\eta}$ at each site. The third term corresponds to the Rashba
SOI of the amplitude $\alpha_{\eta}$. The last two terms represent the intrawire
and interwire pair potential with amplitude $\Delta_{\eta}$ and $\Delta_c$,
respectively. 
To simplify analytical calculations, we focus on the case of $\Delta_1=\Delta_{\bar{1}}$ throughout this paper. Using translational
invariance along $x$-direction, we introduce the number of unit cells $N_x$ and
momentum in $x$-direction $k_x$, and Fourier transform the operators as $%
c_{i\sigma\eta}=\frac{1}{\sqrt{N_x}}\sum_{k_x}e^{ik_xa_x}c_{k_x\sigma\eta}$.
The Hamiltonian can be rewritten in the momentum representation in the basis composed of $c_{k_x}=(c_{k_x%
\uparrow1},c_{k_x\downarrow1},c_{k_x\uparrow\bar{1}},c_{k_x\downarrow\bar{1}%
},c^{\dagger}_{-k_x\uparrow1},c^{\dagger}_{-k_x\downarrow1},c^{%
\dagger}_{-k_x\uparrow\bar{1}},c^{\dagger}_{-k_x\downarrow\bar{1}})^T$ as
\begin{eqnarray}
\mathcal{H}(k_x)&=&\Big[\frac{\xi_{k_x1}}{2}\tau_z+\alpha_1\sin (k_x
a_x)\tau_zs_y\Big](1+\eta_z)  \nonumber \\
&+&\Big[\frac{\xi_{k_x\bar{1}}}{2}\tau_z+\alpha_{\bar{1}}\sin
(k_xa_x)\tau_zs_y\Big](1-\eta_z)  \nonumber \\
&-&\Delta_{1}\tau_ys_y-\Delta_c\tau_ys_y\eta_x  \label{model2}
\end{eqnarray}
Here, $s_{x,y,z}, \tau_{x,y,z},$ and $\eta_{x,y,z}$ are Pauli matrices acting
on the degree of freedom of spin, particle-hole, and chain, respectively. We
define $\xi_{k_x\eta}$ as $\xi_{k_x\eta}=-2t\cos (k_xa_x)-\mu_{\eta}$.

By linearizing the energy spectrum, \cite{JK2012_5}
the authors in the work of Ref. %
[\onlinecite{Interpair}] found that when $\Delta_c>\Delta_{1}$, time reversal
invariant topological superconductor is realized. Here, we study this setup
by tight-binding analysis to find whether time reversal invariant
topological superconductor is obtained beyond the linearized approximation
of the energy dispersion for arbitrary chemical potential.

Below, we calculate winding number to judge whether the system is
topologically non-trivial. With an appropriate choice of the unitary
transformation $V$, the Hamiltonian $H$ can be decomposed into two segments
that are ``time reversal partners" which are distinguished by index $l=1,2$. Then each sector is brought into
off-diagonal form: 
\begin{equation}
V^{\dagger }\mathcal{H}(k_{x})V=\left(
\begin{array}{cc}
\mathcal{H}_{1 }(k_{x}) &0\\
0& \mathcal{H}_{2 }(k_{x})%
\end{array}%
\right) ,
\end{equation}%
\begin{equation}
W^{\dagger }\mathcal{H}_{l}(k_{x})W=\left(
\begin{array}{cc}
0 & h_{l}(k_{x}) \\
h_{l }^{\dagger }(k_{x}) &0%
\end{array}%
\right) ,
\end{equation}%
with the appropriate $8\times 8$ and $4\times 4$ matrices $V$ and $W$ (see
Appendix \ref{Winding number} for more explicit forms). Below, we focus on $%
h_{1}(k_{x})$, however, we note that the topological condition
remains the same if we consider $h_{2}(k_{x})$. The topological
properties can be studied by evaluating the trajectory of determinant of $%
h_{1}(k_{x})$ \cite{wen} as a function of $k_{x}$ for $k_{x}a_x\in
\lbrack -\pi ,\pi )$,
\begin{align}
\text{det}[h_{1}(k_{x})] &=(\xi _{k_{x}1}-2\alpha _{1}\sin
k_{x}a_{x}-i\Delta _{1})  \nonumber \\
&\times (\xi _{k_{x}\bar{1}}-2\alpha _{\bar{1}}\sin k_{x}a_{x}-i\Delta
_{1})+\Delta _{c}^{2}.  \label{det}
\end{align}%
The system is topologically non-trivial when the trajectory of $\det
[h_{1}(k_{x})]$ wraps around the origin in complex space, that is,
the non-zero winding number indicates a non-trivial topological state.
To simplify calculations, it is useful to introduce symmetric and
antisymmetric parameters,
\begin{eqnarray}
\alpha _{s/a} &=&(\alpha _{1}\pm \alpha _{\bar{1}})/2,  \nonumber \\
\mu _{s/a} &=&(\mu _{1}\pm \mu _{\bar{1}})/2.
\end{eqnarray}%
The real and imaginary parts of det$[h_{1}(k_{x})]$ are rewritten as
\begin{align}
\lambda (k_{x})&\equiv \text{Re}\bigl[\text{det}[h_{1}(k_{x})]\bigr]%
=-(\mu _{a}+2\alpha _{a}\sin k_{x}a_{x})^{2} +\Delta _{c}^{2}\nonumber \\
&\vspace{20 pt} -\Delta _{1}^{2}+(2t\cos k_{x}a_{x}+\mu _{s}+2\alpha
_{s}\sin k_{x}a_{x})^{2},  \label{re2} \\
\varepsilon (k_{x})&\equiv \text{Im}\bigl[\text{det}[h_{1 }(k_{x})]%
\bigr]  \nonumber \\
&\vspace{20 pt}=2\Delta _{1}(2t\cos k_{x}a_{x}+2\alpha _{s}\sin k_{x}a_{x}+\mu _{s}).
\label{im2}
\end{align}%
Here, we assume $\sqrt{4t^2+4\alpha_s^2}<|\mu_s|$ such that $\varepsilon(k_{x})$ becomes zero at certain values of $k_x$.
Based on the spirit of \textquotedblleft weak pairing limit", \cite%
{Msato,Qi2010}
we discuss the winding number. \cite{wen} The winding
number is equivalent to the number of wrapping of the normalized vector
\begin{equation}
(d_{1},d_{2})\equiv \left( \frac{\lambda (k_{x})}{\sqrt{\varepsilon
^{2}(k_{x})+\lambda ^{2}(k_{x})}},\frac{\varepsilon (k_{x})}{\sqrt{%
\varepsilon ^{2}(k_{x})+\lambda ^{2}(k_{x})}}\right)
\end{equation}%
around the origin in complex space as $k_{x}$ changes from $k_{x}a_x=-\pi $ to $k_{x}a_x=\pi $. We scale $\lambda (k_{x})$ to $\lambda (k_{x})\rightarrow
a\lambda (k_{x})$ by introducing a parameter $a$, and continuously change $a$
from $a=1$ to a small non-zero value. Importantly, upon this change, the
winding number remains the same. In the case of $\varepsilon (k_{x})\neq 0$,
$(d_{1},d_{2})$ is approximated as $(0, {\rm sgn}\ \varepsilon (k_{x}) )\equiv (0,\pm1)$ for a sufficiently small $a$ and stays constant.
However, in the vicinity of Fermi momenta $k_{F,j}$, which we define as two solutions of $\varepsilon (k_{F,j})=0$  such that $k_{F,1}<k_{F,2}$, the $(d_{1},d_{2})$ vector could wind around the origin.  Close to these momenta, $\varepsilon (k_{x})$ can be expanded as
\begin{equation}
\varepsilon (k_{x})=[\partial _{k_{x}}\varepsilon ](k_{x}-k_{F,j})+\cdots
\end{equation}%
At around the momentum $k_{F,j}$, the vector $(d_{1},d_{2})$ reads
\begin{eqnarray}
d_{1} &=&\frac{a\lambda (k_{x})}{\sqrt{(\partial _{k_{x}}\varepsilon
)^{2}(k_{x}-k_{F,j})^{2}+a^{2}\lambda ^{2}(k_{x})}},  \nonumber \\
d_{2} &=&\frac{[\partial _{k_{x}}\varepsilon ][(k_{x}-k_{F,j})]}{\sqrt{%
(\partial _{k_{x}}\varepsilon )^{2}(k_{x}-k_{F,j})^{2}+a^{2}\lambda
^{2}(k_{x})}}.
\end{eqnarray}%
Now we consider the trajectory spanned the the vector $(d_{1},d_{2})$ as the momentum changes from $k_{x}a_x=-\pi $
to $k_{x}a_x=\pi $, see also Fig. \ref{fig1}b.
As we explained above, only parts of the trajectory close to
$k_{x}=k_{F,1}$ and $%
k_{x}=k_{F,2}$ contribute to the winding number. Thus, we
only focus on the trajectory around these momenta. If we consider $k_{x}$
changing as
\begin{eqnarray}
(k_{x} &<&k_{F,1})\rightarrow (k_{x}=k_{F,1})\rightarrow (k_{F,1}<k_{x}<k_{F,2}) \notag\\
&\rightarrow& (k_{x}=k_{F,2})\rightarrow (k_{x}>k_{F,2})\notag,
\end{eqnarray}%
the vector $(d_{1},d_{2})$ changes accordingly  as
\begin{eqnarray}
(0, \text{sgn}[\varepsilon (k_{x} <k_{F,1}) ]) \rightarrow (\text{sgn}%
[\lambda (k_{F,1})],0)\rightarrow (0,\text{sgn}[\varepsilon
]) \notag\\
\rightarrow (\text{sgn}%
[\lambda (k_{F,2})],0)\rightarrow (0,\text{sgn}[\varepsilon (k_{x} >k_{F,2}) ])\notag.
\end{eqnarray}%
If $\text{sgn}[\lambda (k_{F,1})]=-%
\text{sgn}[\lambda (k_{F,2})]$, the trajectory of the vector $(d_{1},d_{2})$ wraps
around the origin, see Fig.~\ref{fig1}(b), indicating a topologically non-trivial
state. On the other hand, if $\text{sgn}[\lambda (k_{F,1})]=\text{sgn}%
[\lambda (k_{F,2})]$, the trajectory does not wind around the origin, see Fig.~\ref%
{fig1}(c), and the system is in the trivial state. Therefore, we can define the topological condition as follows,
\begin{equation}
(-1)^{\nu }=\prod_{\substack{ k_{x}=k_{F,1},k_{F,2}}}\text{sgn}[\Delta
_{c}^{2}-\Delta _{1}^{2}-(\mu _{a}+2\alpha _{a}\sin k_{x}a_{x})^{2}],
\label{condition}
\end{equation}%
where for $\nu =1(0)$ the system is non-trivial (trivial) topological state.

\begin{figure}[b]
\begin{center}
\includegraphics[width=80mm]{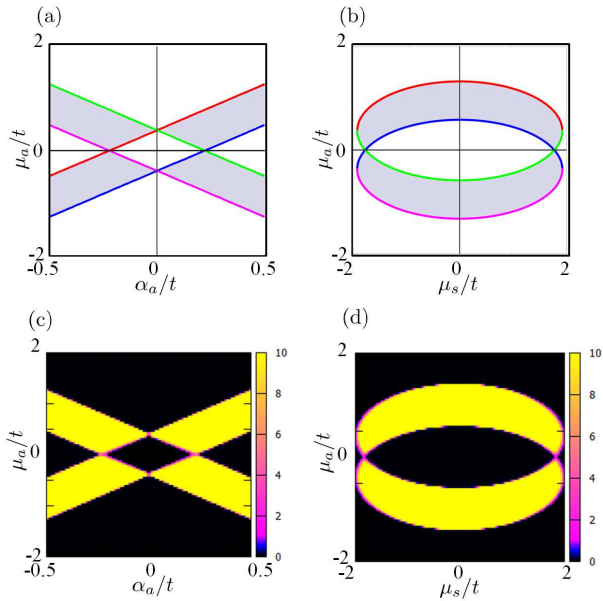}
\end{center}
\caption{(a)[(b)]Topological phase diagram of DQW model as function of $\alpha_a$ and $\mu_a$ ($\mu_s$ and $\mu_a$) with $\Delta_c/t=0.4$, $\Delta_1/t=0.1$, and $\mu_s/t=-1$ ($\Delta_c/t=0.4$, $\Delta_1/t=0.1$, and $\alpha_a/t=0.5$ ). The light gray area represents topologically non-trivial phase. The red, green, blue, and magenta lines correspond to $\mu_a=2\alpha_a\mu_0+\sqrt{\Delta_c^2-\Delta_1^2}, \mu_a=-2\alpha_a\mu_0+\sqrt{\Delta_c^2-\Delta_1^2}, \mu_a=2\alpha_a\mu_0-\sqrt{\Delta_c^2-\Delta_1^2},$ and $\mu_a=-2\alpha_a\mu_0-\sqrt{\Delta_c^2-\Delta_1^2}$, respectively. (c)[(d)] The intensity plot of LDOS of zero energy on the edge of DQW model found by recursive Green's function technique as function of as function of $\alpha_a$ and $\mu_a$ ($\mu_s$ and $\mu_a$) with $\Delta_c/t=0.4$, $\Delta_1/t=0.1$, and $\mu_s/t=-1$ ($\Delta_c/t=0.4$, $\Delta_1/t=0.1$, and $\alpha_a/t=0.5$ ). We set length of the system long enough (4000 sites) so that overlapping between zero energy states on the both edges is negligible. \cite{JK2013_3}
Note that non-trivial area in (a)/(b) well corresponds to that of large amplitude of LDOS in (c)/(d).
}
\label{fig2}
\end{figure}

First, we note from Eq. (\ref{condition}) that the condition $\Delta_c>\Delta_1$ has to be satisfied to
realize topologically non-trivial state otherwise the product in Eq. (\ref%
{condition}) is always positive and the system is in the trivial state. Also, $\alpha_a$ should not be non-zero
because only the term $\alpha_a\sin k_x a_x$ can produce the sign change of $%
[\Delta^2_c-\Delta_1^2-(\mu_a+2\alpha_a\sin k_x a_x)^2]$ which leads minus
sign of the product. Thus, the antisymmetric SOI is crucial for inducing the
topologically non-trivial state.
Otherwise, the SOI could be gauged away as was noted before  in Ref. [\onlinecite{Interpair,Interpair2}].

To get the explicit phase diagram for the system, we consider a simplified case by setting $\alpha_s=0$. In this case, the Fermi points $k_{F,1}$ and $k_{F,2}$  are defined via
\begin{equation}
\varepsilon(k_x)=-2t\cos k_x a_x-\mu_s=0,
\end{equation}
and are given by
\begin{eqnarray}
k_{F,1}a_x=-\arcsin[\mu_0], \
k_{F,2}a_x =\arcsin[\mu_0].
\end{eqnarray}
with $\mu_0\equiv\sqrt{1-(\mu_s/2t)^2}$ for $|\mu_s|<2t$. 
According to Eq.  (\ref{condition}),
\begin{eqnarray}
(-1)^{\nu }={\rm sgn}\Bigl[\Delta_c^2-\Delta_1^2-(\mu_a+2\alpha_a\mu_0)^{2}]\notag\\
\times{\rm sgn}\Bigl[\Delta_c^2-\Delta_1^2-(\mu_a-2\alpha_a\mu_0)^{2}\Bigr].\label{sign}
\end{eqnarray}
has to result in $\nu=-1$ to realize the topological phase.
For $\alpha_a\mu_a>0$,
Eq. (\ref{sign}) is equivalent to
\begin{equation}
 |\mu_a+2\alpha_a\mu_0|>\sqrt{\Delta_c^2-\Delta_1^2}> |\mu_a-2\alpha_a\mu_0| \label{ine}
\end{equation}
%
%
have to be satisfied in order to get non-trivial state.
In Fig. \ref{fig2}(a) [\ref{fig2}(b)], we show the phase
diagram of the DQW system  as  a
function of $\alpha_a$ and $\mu_a$ ($\mu_s$ and $\mu_a$) with fixed $\Delta_c$, $\Delta_1$, and $\mu_s$ ($\Delta_c$, $\Delta_1$, and $\alpha_a$). 
The obtained results are in a good agreement with the results obtained in the tight-binding framework, as the non-zero LDOS at zero energy on edge of the DQW system indicate, see Fig. \ref{fig2}(c) and 2(d). 

To summarize this section, two conditions, $\Delta_c>\Delta_1$ and $%
\alpha_a \neq 0$, are necessarily but not sufficient to generate time reversal invariant topological superconductivity. As we show in the next section, the presence of zero energy state also changes the dominant  symmetries of Cooper pairs at the end of the system.


\section{Cooper pair symmetry}

\label{sec3} In this section, we study symmetries of Cooper pair in the
model of double Rashba QW system coupled to an $s$-wave superconductor. In addition to the standard symmetries of Cooper pair, where frequency, spin, and parity are taken
into account, we should include one more spatial degree of freedom connecting to two QWs. We call the pair amplitude to be wire-odd (wire-even) if it
picks up negative sign (remains the same) by the exchange of the wire index. Due
to this additional degree of freedom, there are now eight classes of Cooper pair
with different symmetries that are consistent with Fermi-Dirac statistics as
summarized  in Table \ref{table}.
These classes are i)
even-frequency spin-singlet even-parity even-wire (ESEE), ii) even-frequency
spin-singlet odd-parity odd-wire (ESOO), iii) even-frequency spin-triplet
odd-parity even-wire (ETOE), iv) even-frequency spin-triplet even-parity
odd-wire (ETEO), v) odd-frequency spin-singlet odd-parity even-wire (OSOE),
vi) odd-frequency spin-singlet even-parity odd-wire (OSEO), vii)
odd-frequency spin-triplet even-parity even-wire (OTEE), and viii)
odd-frequency spin-triplet odd-parity odd-wire (OTOO).

\begin{table}[b]
\begin{tabular}{c||c|c|c|c||c}
\hline
& Frequency & Spin & Parity & Wire & Total \\ \hline\hline
ESEE & +(Even) & -(Singlet) & +(Even) & +(Even) & -(Odd) \\ \hline
ESOO & +(Even) & -(Singlet) & -(Odd) & -(Odd) & -(Odd) \\ \hline
ETOE & +(Even) & +(Triplet) & -(Odd) & +(Even) & -(Odd) \\ \hline
ETEO & +(Even) & +(Triplet) & +(Even) & -(Odd) & -(Odd) \\ \hline
OSOE & -(Odd) & -(Singlet) & -(Odd) & +(Even) & -(Odd) \\ \hline
OSEO & -(Odd) & -(Singlet) & +(Even) & -(Odd) & -(Odd) \\ \hline
OTEE & -(Odd) & +(Triplet) & +(Even) & +(Even) & -(Odd) \\ \hline
OTOO & -(Odd) & +(Triplet) & -(Odd) & -(Odd) & -(Odd) \\ \hline
\end{tabular}%
\caption{All possible symmetries of Cooper pairs occurring in the double QW system.}
\label{table}
\end{table}

The odd-frequency pairings combined with  even-wire symmetry cases, v) and vii), have been studied in previous work dedicated to ferromagnet junctions, unconventional superconductor
junctions, and non-uniform superconducting sytems. \cite%
{Efetov1,Efetov2,Eschrig2003,Asano2007PRL,fominov1,Braude,Asano2007PRB,Yokoyama1,Eschrig2008,Eschrig2015,Keizer,Sosnin,Birge,Robinson,Sprungmann,Anwar,odd1,odd3,odd3b,Eschrig2007,tanaka12}
The odd-frequency pairings combined  with odd-wire symmetry, vi) and viii), have been
discussed in bulk multi band (orbital) systems \cite%
{Balatsky2,Balatsky2015,asano2015} and two-channel Kondo lattice model. \cite%
{Hoshino2015} In the present model, we consider both two effects.

First, we discuss how to evaluate pairing amplitudes. Using Eq. (\ref{model}), we
define Matsubara Green's function as follows:
\begin{equation}
G_M(\omega_n,j,j^{\prime},\sigma,\sigma^{\prime},\eta,\eta^{\prime})=\Bigl(%
\frac{1}{i\omega_n-H}\Bigr)_{j,j^{\prime},\sigma,\sigma^{\prime},\eta,\eta^{%
\prime}}  \label{matsubara}
\end{equation}
with Matsubara frequency which is set to be $\omega_n/t=0.01$ throughout
the paper without loss of generality. Introducing matrices $G(\omega_n,j,j^{\prime},\sigma,\sigma^{\prime},%
\eta,\eta^{\prime})$ and $F(\omega_n,j,j^{\prime},\sigma,\sigma^{\prime},%
\eta,\eta^{\prime})$, we rewrite Eq. (\ref{matsubara}) as
\begin{equation}
G_M=\left(
\begin{array}{cc}
G & F \\
\tilde{F} & \tilde{G}%
\end{array}
\right),  \label{green}
\end{equation}
where $G_M$ is divided into four sectors in the particle-hole space. We focus on
$F(\omega_n,j,j^{\prime},\sigma,\sigma^{\prime},\eta,\eta^{\prime})$ to
analyze symmetries of Cooper pairs. First, we focus on the frequency dependence. Introducing $J\equiv(j,j^{\prime},\sigma,%
\sigma^{\prime},\eta,\eta^{\prime}$), we define  $F_{\text{O}}(J)$ and $F_{%
\text{E}}(J)$ as follows
\begin{equation}
F_{\mathfrak{A}}(J)=\frac{F(\omega_n;J)+\text{sgn}[\mathfrak{A}%
]F(-\omega_n;J)}{2}.  \label{FEO}
\end{equation}
Here, we define $\mathfrak{A}=$E, O with the convention that  sgn$[\text{E}/\text{O}]=\pm1$. Then, using Eq.
(\ref{FEO}) and defining $K$ as $K\equiv(\sigma,\sigma^{\prime},\eta,\eta^{%
\prime})$, we introduce $F_{\text{OO}}(j;K), F_{\text{OE}}(j;K), F_{\text{EO}}(j;K),$ and
$F_{\text{EE}}(j;K)$ classified  also by the spatial symmetry inside the QW as 
\begin{eqnarray}
F_{\mathfrak{A}\mathfrak{B}}(j;K)&=&
\frac{1}{2}\Bigl\{F_{\mathfrak{A}}(j+\frac{1-\text{sgn}[\mathfrak{B}]}{2},j)
\\
&+&\text{sgn}[\mathfrak{B}]F_{\mathfrak{A}}(j,j+\frac{1-\text{sgn}[\mathfrak{%
B}]}{2})\Bigr\} ,   \nonumber
\end{eqnarray}
with $\mathfrak{B}=$E, O. 
Further, by denoting $L$ as $L\equiv(\sigma,\sigma^{\prime})$, we can address the spacial symmetry between QWs  defined by $%
F_{\mathfrak{A}\mathfrak{B}\text{O}}(j;L)$, $F_{\mathfrak{A}\mathfrak{B}%
\text{E}}(j;L)$ as
\begin{equation}
F_{{\mathfrak{A}\mathfrak{B}\mathfrak{C}}}(j;L)=\frac{F_{\mathfrak{A}%
\mathfrak{B}}(j;L;1,\bar{1})+\text{sgn}[\mathfrak{C}]F_{\mathfrak{A}%
\mathfrak{B}}(j;L;\bar{1},1)}{2},
\end{equation}
with $\mathfrak{C}=$E, O. Finally, by addressing also the spin degree of freedom, we get eight classes of pair amplitude
which are given by
\begin{eqnarray}
F^{\uparrow\downarrow}_{\mathfrak{A}\text{T}\mathfrak{B}\mathfrak{C}}(j)=%
\frac{F_{\mathfrak{A}\mathfrak{B}\mathfrak{C}}(j;\uparrow,\downarrow)+F_{%
\mathfrak{A}\mathfrak{B}\mathfrak{C}}(j;\downarrow,\uparrow)}{2}  \label{f}
\\
F_{\mathfrak{A}\text{S}\mathfrak{B}\mathfrak{C}}(j)=\frac{F_{\mathfrak{A}%
\mathfrak{B}\mathfrak{C}}(j;\uparrow,\downarrow)-F_{\mathfrak{A}\mathfrak{B}%
\mathfrak{C}}(j;\downarrow,\uparrow)}{2},
\end{eqnarray}
where the indices $\mathfrak{A}, \mathfrak{B},$ and $\mathfrak{C}$ take the values either E or O.
The combination of $\mathfrak{A}, \mathfrak{B},$ and $\mathfrak{C}$ has to
satisfy
\begin{equation}
\text{sgn}[\mathfrak{A}]\text{sgn}[\mathfrak{B}]\text{sgn}[\mathfrak{C}]=-1
(+1)
\end{equation}
for spin-triplet (spin-singlet) pairing. As for $\uparrow\uparrow$ and $%
\downarrow\downarrow$ spin-triplet components, the corresponding pair
amplitudes are
\begin{eqnarray}
F^{\uparrow\uparrow}_{\mathfrak{A}\text{T}\mathfrak{B}\mathfrak{C}}(j)=F_{%
\mathfrak{A}\text{T}\mathfrak{B}\mathfrak{C}}(j,\uparrow,\uparrow) \\
F^{\downarrow\downarrow}_{\mathfrak{A}\text{T}\mathfrak{B}\mathfrak{C}%
}(j)=F_{\mathfrak{A}\text{T}\mathfrak{B}\mathfrak{C}}(j,\downarrow,%
\downarrow) .  \label{f2}
\end{eqnarray}
We emphasize that due to the time reversal symmetry present in the system,  the  $\uparrow\downarrow$ components of spin-triplet are absent.
Indeed, using
the definition of anomalous Green function
\begin{eqnarray}
&&F^{\dag}(\omega_n,j,j^{\prime},\sigma,\sigma^{\prime},\eta,\eta^{\prime})\\
&&\hspace{40pt}\equiv\int_0^{\beta}d \tau e^{i\omega_n\tau}\left\langle
c^{\dag}(\tau)_{j,\sigma,\eta}c^{\dag}(0)_{j^{\prime},\sigma
^{\prime},\eta^{\prime}}\right\rangle, \nonumber \\
&&F(\omega_n,j,j^{\prime},\sigma,\sigma^{\prime},\eta,\eta^{\prime})
\\
&&\hspace{40pt}\equiv\int_0^{\beta}d \tau e^{i\omega_n\tau}\left\langle
c(\tau)_{j,\sigma,\eta}c(0)_{j^{\prime},\sigma
^{\prime},\eta^{\prime}}\right\rangle \nonumber
\end{eqnarray}
with imaginary time $\tau $ and inverse temperature $\beta =1/k_{B}T$, as
well as the fact that
\begin{equation}
c(\tau )_{j,\uparrow ,\eta }\overset{\mathfrak{T}}{\rightarrow }-c(\tau
)_{j,\downarrow ,\eta }
\end{equation}%
\begin{equation}
c(\tau )_{j,\downarrow,\eta }\overset{\mathfrak{T}}{\rightarrow }c(\tau
)_{j,\uparrow ,\eta }
\end{equation}%
under time reversal operation $\mathfrak{T}$, we obtain
\begin{eqnarray}
\mathfrak{T}[F(M;\uparrow,\downarrow;N)+F(M;\downarrow ,\uparrow ;N)]%
\mathfrak{T}^{-1}  \notag \\
=-[F(M' \downarrow ,\uparrow ;N)+F(M' \uparrow ,\downarrow ;N)]
\end{eqnarray}%
for the time-reversal invariant system. 
Using the fact that Green's function
in Eq. (\ref{green}) is invariant under time reversal operation 
we conclude that the  $\uparrow\downarrow$ components of spin-triplet are absent.
Above we used notations $M\equiv (\omega _{n},j,j^{\prime })$, $M^{\prime}\equiv (-\omega _{n},j,j^{\prime })$, and
$N\equiv
(\eta ,\eta ^{\prime })$. 
%
By similar argument, we show that $\uparrow \uparrow $ and $\downarrow
\downarrow $ components are equal.

The Eqs. (\ref{f})-(\ref{f2})  describe all possible types of pair amplitudes that are also represented in Table \ref{table}. Before we begin with numerical
calculations, we clarify general properties of the Hamiltonian and resulting
symmetries of pairing amplitudes. First, we consider the infinite DQWs model. We start from the case with $\mu_{a}=0$, $\alpha_{s}=0$
and $\alpha_{a}=0$. In this case, only the ESEE pairing is present. This is the
standard pairing for $\Delta_{1}$ and
$\Delta_{c}$ originating from an $s$-wave superconductor without any symmetry breaking in double wire and spin-rotational spaces as well as without breaking of the translational invarience. As shown in Table \ref{table2}, by breaking these symmetries, seven additional  types of superconducting pairing are
induced. In case (1), we break only the double wire
symmetry by adding non-zero $\mu_{a}$.  Now, the double-wire-even and double-wire-odd pairings can mix without changing symmetries in the spin space and without breaking the translational invarience. To be consistent with Fermi-Dirac statistics,
the parity in the frequency space should be switched, thus the OSEO pairing is induced. In case (2), only $\alpha_{s}$ is chosen to be nonzero, thus, the spin rotational symmetry
and spatial parity inside wire are broken at the same time. Then, the ETOE
pairing is induced without generating odd-frequency pairing as was shown before in
non-centrosymmetric superconductors with Rashba SOI. \cite{Yada2009} Next, in the case
(3), the induced symmetries can be understood by combining the results of the cases
(1) and (2). In addition to the ESEE pairing, the OTOO, OSEO, ETOE pairings are induced by the coexistence of $\mu_{a}$ and $\alpha_{s}$.  We note that the
presence of asymmetric Rashba coupling $\alpha_{a}$  also corresponds to the case (3), as
it breaks the double wire symmetry and plays a role similar to $\alpha_{s}$ in the spin space.

Next, we consider systems, in which the translation invariance is also broken, for example, if some parameters are non-uniform or the system is finite.
Thus, the parity mixing can occur by reversing the parity corresponding to
 frequency \cite{odd3,odd3b,tanaka12} to be consistent with Fermi-Dirac
statistics. We first consider the case of $\mu_{s}=0$ and $\alpha_{s}=0$.
In case (4), the breakdown of the parity inside a wire induces the OSOE pairing.
This result is consistent with preexisting results in non-uniform
superconducting systems. \cite{odd3,odd3b,Yokoyamavortex,tanaka12} The
obtained induced pairings in case (5) can be understood by combining the
results of cases (1) and (4). The ESOO pairing is induced by the breakdown of the translational invarience and the double wire symmetry. Also the results in the case (6) can be understood by combining the results in the cases (2) and (4).  The OTEE pairing is induced from the ETOE pairing by breaking of the translation invariance. The most interesting situation
is the case (7). The OSEO, ETOE and OTOO pairings stem from results in the case (3).
Similar to the case (4), the OSOE,  ESOO, OTEE and ETEO pairings are generated by the spatial parity mixing due to the fact that the translation invariance is broken.


\begin{table}[t]
\scalebox{0.82}{
\begin{tabular}{c|ccc|c}\hline
&&Broken symmetry&&Pairing symmetry\\\hline
&DQW&spin&translation&\\
&$\mu_a$&$\alpha_s$&boundary&\\\hline\hline
(0)&--&--&--&ESEE\\\hline
(1)&$\bigcirc$&--&--&ESEE, OSEO\\\hline
(2)&--&$\bigcirc$&--&ESEE, ETOE\\\hline
(3)&$\bigcirc$&$\bigcirc$&--&ESEE, OSEO, ETOE, OTOO\\\hline
(4)&--&--&$\bigcirc$&ESEE, OSOE\\\hline
(5)&$\bigcirc$&--&$\bigcirc$&ESEE, OSEO, OSOE, ESOO\\\hline
(6)&--&$\bigcirc$&$\bigcirc$&ESEE, ETOE, OSOE, OTEE\\\hline
(7)&$\bigcirc$&$\bigcirc$&$\bigcirc$&ESEE, OSEO, ETOE, OTOO\\
&&&&OSOE, ESOO, OTEE, ETEO\\
\hline
\end{tabular}}
\caption{Eight classes of the possible symmetries of Cooper pairs in the DQW system.
The classes are characterized by broken symmetries. The symmetry between two QWs composing the DQW system could be broken by detuning of the chemical potential. The spin space symmetry could be broken by the SOI. The translational invarience is broken, for example, by boundary conditions. The label $\bigcirc$ (-- ) indicates that the symmetry is broken (preserved).}
\label{table2}
\end{table}

\begin{figure}[h]
\begin{center}
\includegraphics[width=80mm]{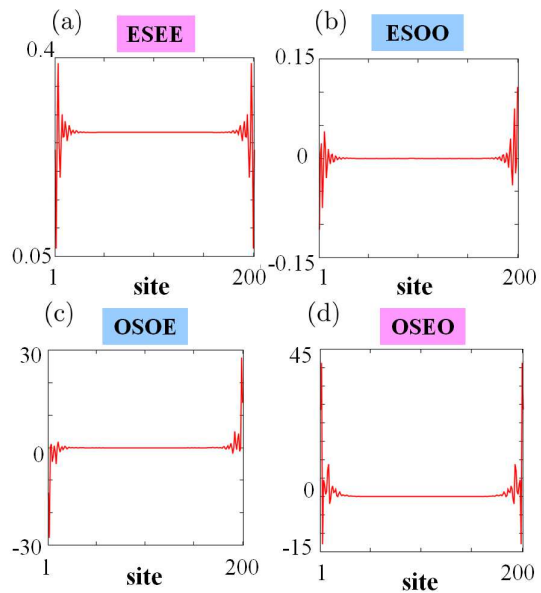}
\caption{The spatial profiles of four singlet pairing amplitudes (denoted above each figure) for the DQW system of the finite size of 200 sites. The parameters are set to be as follows: $\alpha_1/t=-\alpha_{\bar{1}}/t=0.5, \mu_1/t=-2, \mu_{\bar{1}}/t=-1, \Delta_1/t=0.1,$ and $\Delta_c/t=0.4$. We confirm that the odd-frequency pairing amplitudes are strongly enhanced at both edges as panels (c) and (d) show.  }\label{fig3}
\end{center}
\end{figure}
\begin{figure}[h]
\begin{center}
\includegraphics[width=80mm]{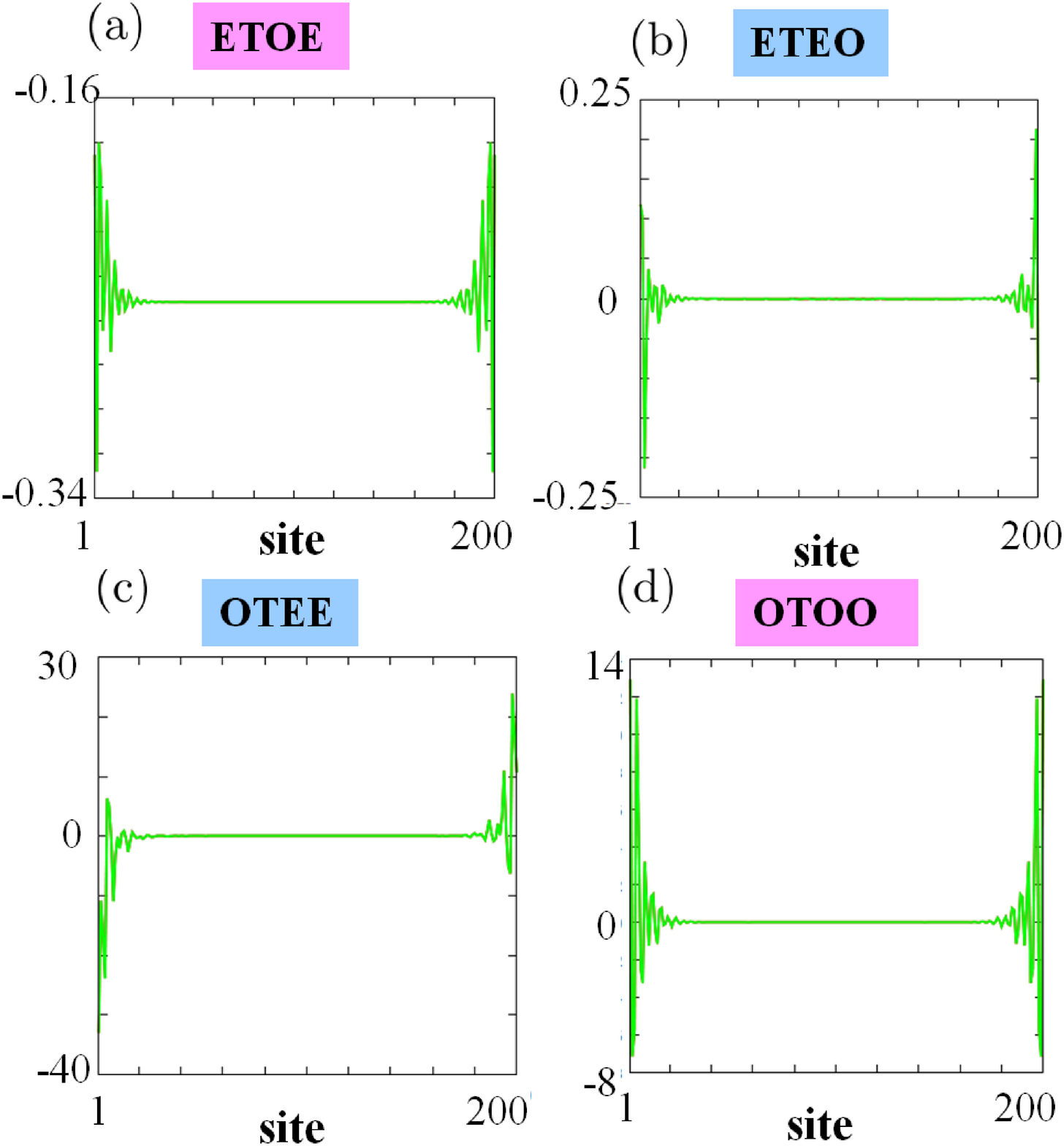}
\caption{The spatial profiles of four triplet pairing amplitudes. The parameters are the same as in Fig. \ref{fig3}.
Again, the odd-frequency pairing amplitudes peak at the edges, where Kramer pairs of  MFs are located, see panels (c) and (d)}\label{fig4}
\end{center}
\end{figure}

We calculate the spatial profile of pairing amplitudes numerically for parameters chosen such that the system is in the topological regime, see Figs. \ref{fig3} and \ref{fig4}.
First, odd-frequency components, $i.e.$ the magnitudes of the pairing amplitudes of
OSEO, OTOO, OSOE, and OTEE, are hugely enhanced at the system edge in consistence with  the existence of zero energy state, i.e., Kramers Majorana fermions, similar to the previous results obtained in unconventional superconductors. \cite{odd1,odd3,odd3b}
Second, in addition to the ESEE pairing, which is the primary symmetry of
the parent system  [see Fig. \ref{fig3}(a)], the ETOE pairing spreads over the system [see Fig. \ref{fig4}(a)]. Although the OTOO and OSEO pairings are possible in the bulk from the discussion of the pairing symmetries (see above  in Table \ref{table2}), their magnitudes are
small. The ESOO and ETEO pairing strengths are small and non-zero only at the system edge
due to the breakdown of translational symmetry. To understand the spatial profile of these pairing amplitudes, it is convenient to  focus on the inversion parity of the pairing amplitudes around the center of the quantum wire. As seen from Figs. \ref{fig3} and \ref{fig4},
the inversion parity is even for the ESEE, OSEO, ETOE, OTOO pairings. These pairings can exist also in the bulk. In contrast to that, the inversion parities of the
OSOE, ESOO, OTEE, ETEO pairings are odd. They are generated due to the breakdown of the
translational invariance and localized at the edges.

\begin{figure}[h]
\begin{center}
\includegraphics[width=80mm]{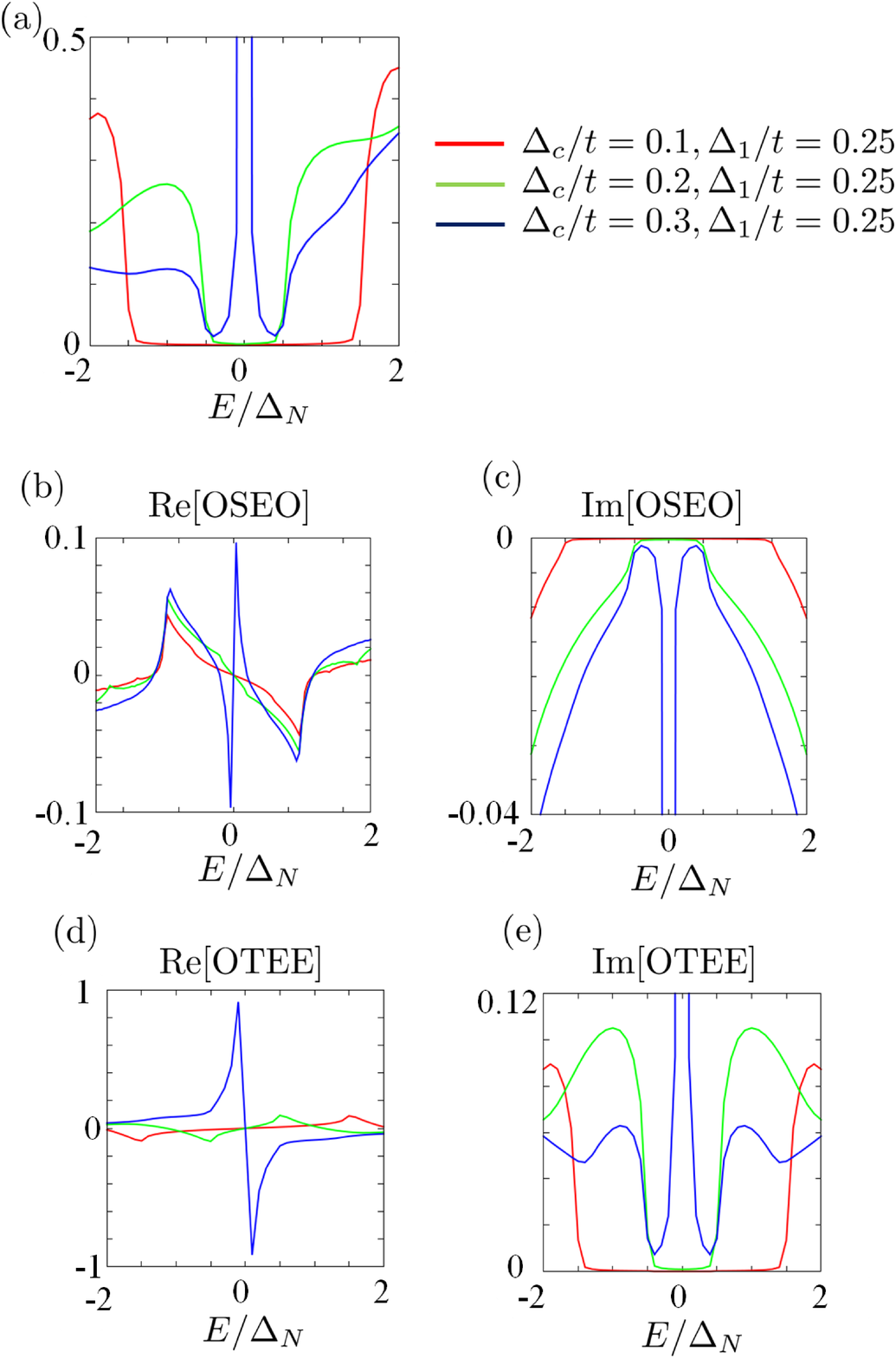}
\caption{(a) The LDOS on the edge of the DQW system of the length of 4000 sites. (b-c)
The real and imaginary parts of the OSEO pairing. (d-e): The real and imaginary parts of the OTEE pairing. Plots are based on recursive the Green function technique.
The parameters are set as follows, $\alpha_1/t=-\alpha_{\bar{1}}/t=0.5$, $\mu_1/t=-2$, and $\mu_{\bar{1}}/t=-1$. 
}
\label{figdos}
\end{center}
\end{figure}

To emphasize the correspondence between zero energy states and odd-frequency pairings  explicitly, we calculated numerically the LDOS at the edge of the DQW system and the pairing amplitudes of OSEO and OTEE (see Fig. \ref{figdos}) for three different cases. 
The pairing amplitudes [see Figs. \ref{figdos}(b)-(e)] change as a function of energy similar to the LDOS [see Fig. \ref{figdos}(a)]. Specifically, when parameters are set to satisfy the topological criterion (blue line in Fig. \ref{figdos}), the real parts of the OSEO and OTEE pairing amplitudes change abruptly around zero energy. Importantly, the imaginary parts of the OSEO and OTEE  pairing amplitudes peak strongly at zero energy, confirming the connection between the presence of the MFs and the odd-frequency pairing.

%

\section{Josephson junction of DQWs}

\label{sec4} In this section, we address Andreev bound states (ABSs) in DQW/DQW
junctions. The ABSs localized between two superconductors
are extensively studied in the literature. The energy of the ABS $E_{b}$ localized between two topologically trivial $s$-wave superconductors  is given by
$E_{b}=\pm \sqrt{1-\sigma _{N}\sin^{2}(\varphi /2)}\Delta _{0}$,
where $\Delta _{0}$, $\varphi $, and $\sigma
_{N}$ is the magnitude of the superconducting pairing potential, the phase difference between superconductors and
the transparency of the junctions, respectively. \cite{FT1991b} On the other hand, the energy of ABS $E_{b}$ between two
one-dimensional topological $p_x$-wave superconductors
is given by $E_{b}=\pm \sqrt{\sigma _{N}}\Delta_{0} \cos \varphi /2 $.\cite{Kwon,Kitaev01}
The similar result is also known from the study of $d$-wave
superconductor junctions. \cite{Tanaka97,kashiwaya00} The present anomalous $\varphi $ dependence of the ABS can be explained by the coupling of MFs on both
sides in Kitaev chain/Kitaev chain Josephson junction system.\cite{Kitaev01}
It also generates $4\pi $ periodicity of the AC Josephson current in the Josephson
junctions based on topological superconductors.
In this section, we calculate the energy of the ABSs
and find anomalous $\sim\pm\sin\varphi$ dependence.
We provide qualitative explanation of this sinusoidal curve by considering  the coupling of the Kramers pair of MFs to an $s$-wave superconductor. We also construct the effective model, which address the $s+p$/$s$-wave superconductor junctions, to explain this phase dependence of the ABS energy.

\begin{figure}[b]
\begin{center}
\includegraphics[width=80mm]{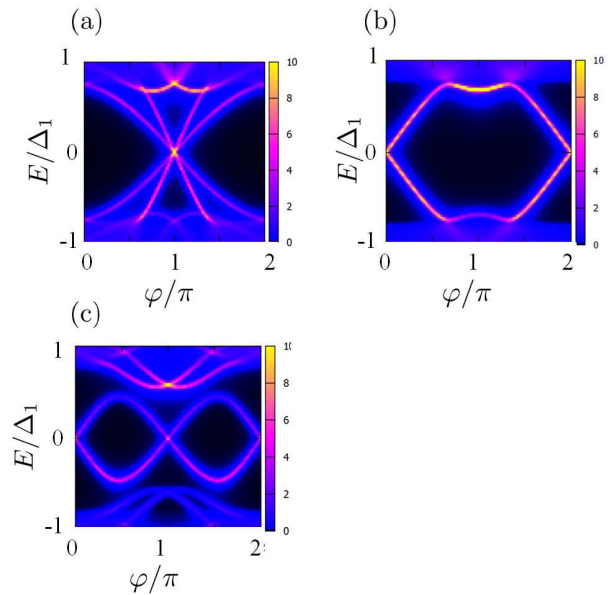}
\end{center}
\caption{(a) The spectrum of the ABSs in the DQWs/normal metal/DQWs junction as a function of the phase difference $\protect\varphi$. The parameters are set to be the same on both sides and correspond to the topologically non-trivial regime: $\alpha_1/t=-\alpha_{\bar{1}}/t=0.5, \mu_1/t=-2, \mu_{\bar{1}}/t=-1, \Delta_1/t=0.1,$ and $\Delta_c/t=0.4$. The ABSs behave like $\sim\pm\cos{\varphi/2}$.
(b) Next, we reverse the signs of the Rashba SOI on the right side, $i.e.$,  $\alpha_1/t=-\alpha_{\bar{1}}/t=-0.5$, other parameters are set to be the same as in the panel (a). The ABSs exhibit the $\pi$-shift and their energy behaves like $\sim\pm\sin{\varphi/2}$.
(c) The right side of the system is brought into the trivial regime with parameters chosen to be $\alpha_1/t=\alpha_{\bar{1}}/t=0, \mu_1/t=-1, \mu_{\bar{1}}/t=-1,$ and $\Delta_1/t=0.1$. The ABSs near Fermi energy level behave like $\sim\pm\sin{\varphi}$.
For all figures the calculation is based on recursive Green's function technique, and we set the length of the right and left superconducting sections (normal metal section) to 4000 (2) lattice sites. The chemical potential in the normal section is set to $\mu_N/t=-1$. }
\label{fig6}
\end{figure}

Making use of the recursive Green's function technique, we can calculate the spectrum and the energy of the ABSs.
First, we focus on the case when both sides of the DQWs are in the non-trivial topological regime. The result is shown in Fig. \ref{fig6}(a). This behavior is similar to the
case of Kitaev chain/Kitaev chain or $p$-wave/$p$-wave junction system which
demonstrates\cite{Kitaev01} $\sim\pm\cos\varphi/2$. By introducing $%
\gamma_{\uparrow}$ and $\gamma_{\downarrow}$ to describe two different MFs building up the Kramers pair, we can understand the curve in Fig. \ref{fig6}(a) by the
coupling of $\gamma_{\uparrow}$ and  $\gamma_{\downarrow}$ on both
sides, analogously to the mentioned above Kitaev chain/Kitaev chain junction.  If the signs of the Rashba SOI is reversed on the right side, the spectrum of the ABSs is trivially shifted by $\pi$, as shown in Fig. \ref{fig6}(b). This is an example of the so-called $0$-$\pi$ transition: by reversing the sign of the Rashba term, the phase of effective $p$-wave superconductor is flipped by $\pi$. This transition has been already discussed in the system of a Rashba quantum wire on a superconductor with applied Zeeman fields, where topological superconductivity of the class D is realized.\cite{0pi,hanson2015,KnaEPJB2015}
Indeed, the observed behavior [see Fig. \ref{fig6}(b)] can
also be explained by the above mentioned transition using the effective model of DQWs discussed below.

Next, we check the most interesting regime in which we set the one side of the junction to be in the topological regime and another to be in the trivial regime [see, Fig. \ref{fig6}(c)].
The energy spectrum of the ABSs follows $\sim\pm\sin\varphi$.
This feature can be explained by the coupling of Kramers pair of MFs and to the $s$%
-wave superconductor as we demonstrate below.

The tunneling Hamiltonian can be written as
\begin{equation}
H_{T}=\sum_{k\sigma }\Bigl(T_{k}\gamma _{\sigma }b_{k\sigma }+%
\text{h.c.}\Bigr)\label{tun},
\end{equation}%
where $T_{k}$ is the tunneling amplitude, $b_{k\sigma }$ is the
annihilation operator acting on the electron on the ride side of the
junction with the spin $\sigma $ and the momentum $k$. The Majorana operator
$\gamma _{\sigma }$ acts on the left side of the junction with the index $\sigma $ used to distinguish between to two MFs building a Kramers pair. We introduce the Bogoliubov transformation as follows
\begin{eqnarray}
b_{k\uparrow } &=&u_{k}\alpha _{k\uparrow }+v^{*}_{k}\alpha _{-k\downarrow
}^{\dagger }  \notag \\
b_{k\downarrow } &=&u_{-k}\alpha _{k\downarrow }-v^{*}_{-k}\alpha _{-k\uparrow
}^{\dagger },
\end{eqnarray}%
where $u_{k}$ and $v_{k}$ are given by
\begin{equation*}
u_{k}=\frac{1}{2}\sqrt{1+\frac{\xi _{k}}{E_{k}}},\ \ v_{k}=\frac{1}{2}\sqrt{%
1-\frac{\xi _{k}}{E_{k}}}\exp(i\varphi),
\end{equation*}%
and $\varphi$ is the macroscopic phase difference between right
and left superconductors.
Here, $\xi _{k}$ is the energy spectrum of the right superconductor in the
normal state and $E_{k}=\sqrt{\xi _{k}^{2}+|\Delta _{k} |^{2}}$ is the
quasi-particle energy spectrum, where $\Delta_{k}$ is the pair potential
in the right side superconductor. Without loss of generality, $u_k$ can be set to be a real number and $v_k$ to be a complex number.
As a result, the tunneling Hamiltonian is
rewritten as
\begin{align}
&H_{T}=\sum_{k}\Bigl\{T_{k}\gamma _{\uparrow }(u_{k}\alpha
_{k\uparrow }+v^{*}_{k}\alpha _{-k\downarrow }^{\dagger })+T_{k}^{\ast
}(u_{k}\alpha _{k\uparrow }^{\dagger }+v_{k}\alpha _{-k\downarrow })\gamma
_{\uparrow }  \notag \\
& +T_{k}\gamma _{\downarrow }(u_{-k}\alpha _{k\downarrow }-v^{*}_{-k}\alpha
_{-k\uparrow }^{\dagger })+T_{k}^{\ast }(u_{-k}\alpha _{k\downarrow
}^{\dagger }-v_{-k}\alpha _{-k\uparrow })\gamma _{\downarrow }\Bigr\}.
\end{align}%
At the next step, we construct the effective Hamiltonian describing the coupling
between two MFs in the second order perturbation theory,
\begin{equation}
H_{\text{MF}}=\left\langle 0\right\vert H_{T}\frac{1}{E-%
H_{0}}H_{T}\left\vert 0\right\rangle   \label{HMF},
\end{equation}%
where $H_0$ represents BdG Hamiltonian {\it without} tunneling (\ref{tun}), $H_0\equiv\sum_{k\sigma}E_k\alpha_{k\sigma}^{\dagger}\alpha_{k\sigma}$,
and $\left\vert 0\right\rangle $ is set to be the ground-state
of the quasi-particles. From Eq. (\ref{HMF}), one finds
\begin{widetext}
\begin{align}
& H_{T}\left\vert 0\right\rangle =\sum_{k}(T_{k}v_{k}\gamma
_{\uparrow }\alpha _{-k\downarrow }^{\dagger }-T_{k}^{\ast }u_{k}\gamma
_{\uparrow }\alpha _{k\uparrow }^{\dagger }-T_{k}v_{-k}\gamma _{\downarrow
}\alpha _{-k\uparrow }^{\dagger }-T_{k}^{\ast }u_{-k}\gamma _{\downarrow
}\alpha _{k\downarrow }^{\dagger })\left\vert 0\right\rangle , \\
& \left\langle 0\right\vert H_{T}=\left\langle 0\right\vert
\sum_{k^{\prime }}(T_{k^{\prime }}^{\ast }v_{k^{\prime }}^{\ast }\alpha
_{-k^{\prime }\downarrow }\gamma _{\uparrow }-T_{k^{\prime }}u_{k^{\prime
}}\alpha _{k^{\prime }\uparrow }\gamma _{\uparrow }-T_{k^{\prime
}}^{\ast }v_{-k^{\prime }\uparrow }^{\ast }\alpha _{-k^{\prime }}\gamma
_{\downarrow }-T_{k^{\prime }}u_{-k^{\prime }}\alpha _{k^{\prime
}\downarrow }^{\dagger }\gamma _{\downarrow }),\\
&H_{\text{MF}}=-\sum_{k}\frac{|T_{k}|^{2}}{E_{k}}\Bigl\{%
2u_{k}^{2}+2|v_{k}|^{2}+\bigl(\frac{\Delta _{k}}{2E_{k}}-\frac{\Delta
_{k}^{\ast }}{2E_{k}}\bigr)\gamma _{\uparrow }\gamma _{\downarrow }-\bigl(%
\frac{\Delta _{k}}{2E_{k}}-\frac{\Delta _{k}^{\ast }}{2E_{k}}\bigr)\gamma
_{\downarrow }\gamma _{\uparrow }\Bigr\}.  \label{HMF2}
\end{align}
\end{widetext}
Here, we used the fact that $u_k=u_{-k}$, $v_k=v^{*}_{-k}$, and $T_k=T^{*}_{-k}$. and also set $E=0$. First, we ignore the first two terms in Eq. (\ref{HMF2})
because they are constant by noting $u_{k}^{2}+|v_{k}|^{2}=1$. If we impose the phase difference between the left and right
superconductors as $\Delta _{k}\rightarrow \bar{\Delta} _{0}e^{i\varphi }$,
$H_{\text{MF}}$ can be written in the
basis of $(\gamma _{\uparrow },\gamma _{\downarrow })^{T}$ as
\begin{equation}
H_{\text{MF}}=\sum_{k}
\bar{\Delta}_{0}\frac{|T_{k}|^{2}}{E_{k}^{2}}\sin
\varphi \left(
\begin{array}{cc}
0 & -i \\
i & 0%
\end{array}%
\right) .
\end{equation}%
with the energies given by $\pm \frac{|T_{k}|^{2}}{E_{k}^{2}}
\bar{\Delta} _{0}\sin \varphi $, which coincides with the dispersion relation obtained for the
ABSs in Fig.~ \ref{fig6}(c).

The characteristic features of the energy dispersion of the ABSs shown in
Fig. \ref{fig6}(c) are also reproduced by the quasi-classical theory based
on the effective model discussed in Ref. [\onlinecite{Fb}]. Following Ref. %
[\onlinecite{Fb}], we treat the terms breaking the
symmetry between QWs as perturbations. Converting the lattice model back to
continuous one, the effective Hamiltonian is written as
\begin{align}
&\mathcal{H}_{qk}=(\frac{\hbar^{2} k_{x}^{2}}{2m}-2t-\mu _{s})\tau _{z}-(\Delta
_{s}+k_{x}s_{y}\Delta _{p})s_{y}\tau _{y}\label{effective2},\\
&\Delta _{s} =\Delta _{1}-\Delta _{c}+\frac{\mu_{a}^{2}+k_{x}^{2}\alpha
_{a}^{2}}{\Delta _{1}+\Delta _{c}},  \label{s-wave} \\
&\Delta _{p} =2\frac{\alpha _{a}\mu _{a}}{\Delta _{1}+\Delta _{c}},
\label{p-wave}
\end{align}%
where $m$ is the effective mass given by $m=\hbar^{2}/2t a_{x}^{2}$.
 As seen from Eq. (\ref%
{p-wave}), if the signs of Rashba SOI are reversed, the phase of effective $p$%
-wave pair potential is shifted by $\pi $, which explains the 0-$\pi $
transition in Fig. \ref{fig6}(b).  For simplicity, we set the $s$%
-wave pair potential to be determined at the Fermi momentum
\begin{eqnarray}
\Delta _{s} &=&\Delta _{1}-\Delta _{c}+\frac{\mu_{a}^{2}+k_{F}^{2}\alpha
_{a}^{2}}{\Delta _{1}+\Delta _{c}}  \notag \\
&=&\Delta _{1}-\Delta _{c}+\frac{\mu_{a}^{2}+2m(2t+\mu _{s})\alpha _{a}^{2}}{%
\Delta _{1}+\Delta _{c}}.  \label{effective}
\end{eqnarray}%
If we set $\Delta _{p}>\Delta _{s}/\sqrt{2m(2t+\mu _{s})}$, time reversal
topological superconductivity is realized.\cite{Fb} The effective
Hamiltonian describes a one-dimensional $s$+$p$-wave superconductor. Generally, it is
known that if $p$-wave component of the pair potential is larger than $s$%
-wave component, topological superconductor hosting edge states is realized.\cite%
{TYBN08}

\begin{figure}[h]
\begin{center}
\includegraphics[width=82mm]{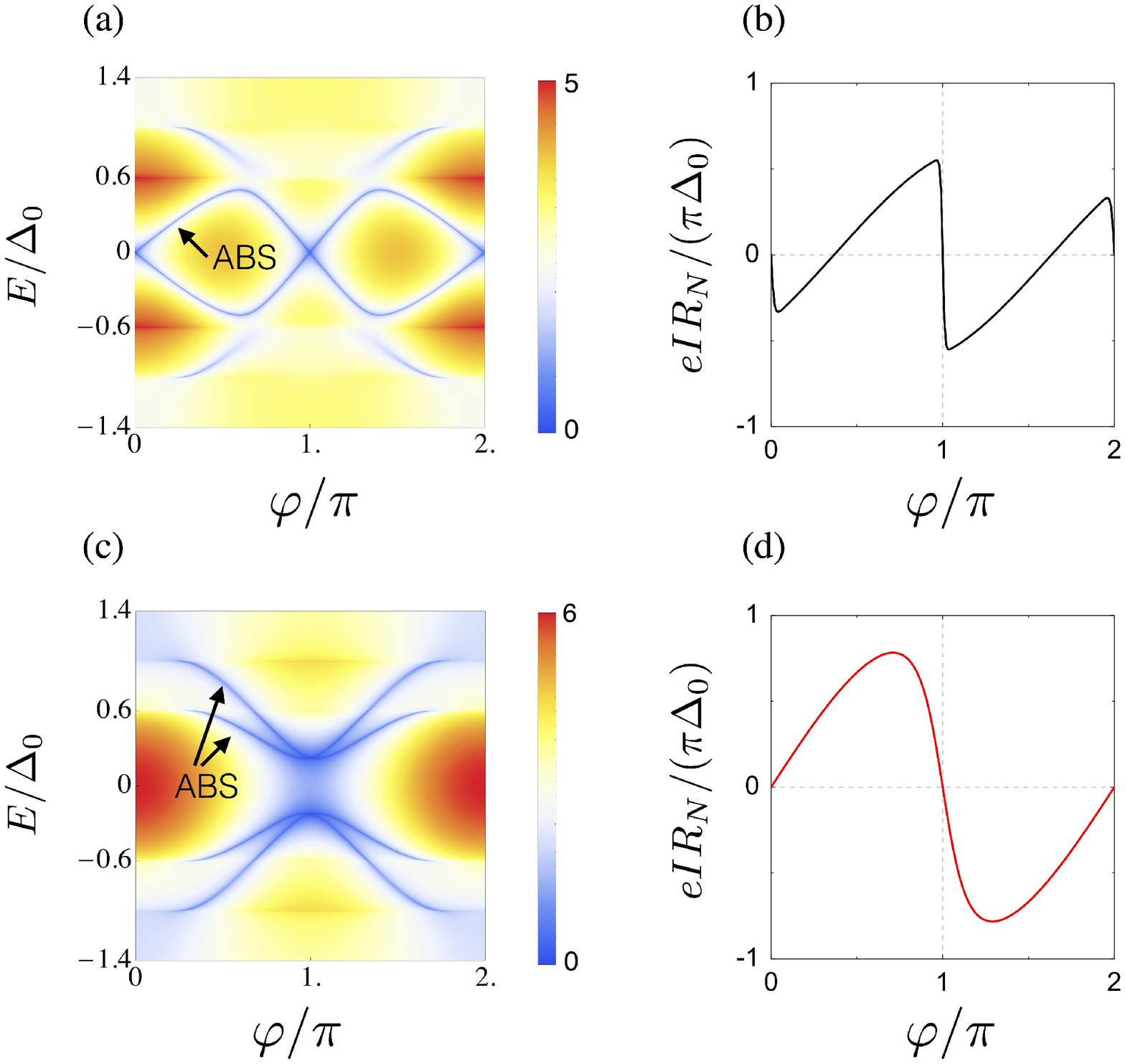}
\end{center}
\caption{Illustration of ABSs and Josephson currents of $s$+$p$-wave/$s$-wave junction produced by continuum model. Panels (a) and (c) 
show $\left\vert \det M\right\vert ^{\frac{1}{4}}$ where the deep blue curve shows the position of ABSs. Panels (b) and (d) show the current-phase relation  normalized to the interface
resistance $R_{N}$ in the normal state.  (a) and (b): $\Delta _{0}/\tilde{\mu}=0.001$, $%
\Delta _{s}/\tilde{\mu}=0.0004$,  $k_{F}\Delta _{p}/\tilde{\mu}=0.001$ and
$\tilde{\mu}=2t+\mu _{s}$. (c) and (d): $\Delta _{0}/\tilde{\mu}=0.001$, $%
\Delta _{s}/\tilde{\mu}=0.001$, and  $k_{F}\Delta _{p}/\tilde{\mu}=0.0004$. We
use $Z=0.5$ for all cases and set the temperature $k_{B}T=0.01\Delta _{0}$ for the panels (b) and (d). Notice that when $p$-wave is dominant, the ABSs near Fermi energy has $\sim \sin \varphi$ feature which corresponds to non-trivial/trivial junction case as shown in Fig. \ref{fig6}(c). 
}
\label{fig7}
\end{figure}

Next, we show that the behavior of ABSs obtained in Fig. \ref{fig6}(c) can be
explained using the effective model of a semi-infinite $s$/$s$+$p$-wave
junction. We consider the pair potential on $s$+$p$-wave side to be given by
Eqs. (\ref{s-wave}) and (\ref{p-wave}), while on the $s$-wave side, it has
the form $\Delta _{s}s_{y}\tau _{y}$ with $\Delta _{s}=\Delta
_{0}e^{i\varphi }$. We assume that $\mu $ and $t$ are the same on
both sections of the junction. However, two sets of values are chosen such that one corresponds to the topological phase and another to the trivial phase. We also take into account the strength of the interface barrier denoted by $Z$ but its value does not influence on the key feature of our results. In order
to calculate ABSs, we seek the solution of $M\hat{y}=0$ where $\hat{y}$ is
the envelope function composed of all the coefficients of the out-going modes (see Appendix \ref{bo-san}).
Then, the ABSs can be obtained from the condition $\det M=0$. To make our
plots more clear, we  show $\left\vert \det M\right\vert ^{%
\frac{1}{4}}$ as a function of energy $E$ and phase difference $\varphi $. The deep blue curve indicated the ABSs in Figs. %
\ref{fig7}(a) and \ref{fig7}(c). In the topological phase,
where the $p$-wave pairing is dominant, we can see that the energy of ABSs is described by $\pm \sin \varphi $ around Fermi energy level,  which is similar to the behaviour observed above
in the lattice model [see Fig. \ref{fig6} (c)]. Thus, the Josephson
current-phase relation is approximated also by non-trivial dependence $\sim \sin 2\varphi $ as
shown in \ref{fig7}(b), apart from standard $2\pi $-periodicity. Importantly, the observed double crossing points of ABSs around Fermi energy, in spite of the fact that the non-zero interface barrier $Z$ is explicitly introduced in the continuum model,
confirm the topological origin of the obtained ABSs (MFs). As for the topologically trivial case in Fig. \ref{fig7}%
(c), where $s$-wave component is dominant, the ABSs behave as $\sim \sqrt{%
1-\sigma _{N}\sin ^{2}\varphi }$. The corresponding current-phase relation has
the standard $\sim \sin \varphi $ feature [see Figs. \ref{fig7}(d)]. It is
also seen that the ABSs are gaped around zero energy due to the presence of the
interface barrier,  which reveals its trivial non-topological nature.

\section{Conclusion}\label{sec5}
In this paper, we have studied the double quantum wire system brought into proximity to an $s$-wave superconductor. As was shown before, the system is in the topological phase if  the induced interwire (crossed Andreev) pairing dominates. We have generalized this topological criterion to account for the detuning of the chemical potential by calculating the winding number. We have also classified symmetry of the superconducting order parameter by focusing on four degrees of freedom, $i.e.$, frequency, spin, the spatial parity inside the QWs, and the spatial parity between the QWs. The magnitude of the odd-frequency pairing is hugely enhanced if the topological superconductivity is realized.
We have also considered the ABSs in the DQW/DQW junctions.
For topological/non-topological junctions, the energy dispersion of the ABSs is proportional to $\sim\pm\sin\varphi$, where $\varphi$ denotes the phase difference between two section of the DQW system.
We have explained this  behavior in terms of the  couplings of Kramers pair of Majorana fermions and spin-singlet $s$-wave Cooper pair.
We have confirmed that
this $\varphi$ dependence can be reproduced using the effective continuum model corresponding to the $s+p$/$s$-wave superconductor junction system.
The odd-frequency pairing and the ABSs in
the topological regime can be detected by tunneling spectroscopy. \cite{TK95,Law09,Yakovenko,Bolech,Tanaka09,Linder10}
However, such calculations are beyond the present work.

\section*{Acknowledgment}
H.E. thanks P. Burset and K. T. Law for useful discussion.
This work was supported by
the Grant-in Aid for Scientific Research on Innovative Areas ``Topological Material Science"
(Grant No. 15H05853),
Grant-in-Aid for Scientific Research B (Grant No. 15H03686),
Grant-in-Aid for Challenging Exploratory Research (Grant No. 15K13498), and
Grant-in-Aid for JSPS Fellows.  We acknowledge support from the Swiss National Science Foundation and the NCCR QSIT.\\
\begin{widetext}
\appendix

\section{Winding number}

\label{Winding number}
In this section of the Appendix, we introduce the winding number to help us to
determine whether the zero energy state found in the main text and in Fig. \ref{fig2} is topologically protected. The procedure is the following. First, we decompose
the Hamiltonian into two sectors that are not coupled to each other. We note that the time reversal partners always belong to the different sectors. Second, we bring the chosen sector to the off-diagonal form and calculate its determinant. By considering how many time the vector composed from the real and imaginary parts of the determinant wraps around the origin in the complex plane as a function of the momentum is nothing but the winding number.


The model Hamiltonian given by Eq. (\ref{model2}) can be rewritten in the
basis composed of $c_{k_{x}}^{\prime }=(c_{k_{x}1\uparrow },c_{-k_{x}1\uparrow
}^{\dagger },c_{k_{x}\bar{1}\uparrow },c_{-k_{x}\bar{1}\uparrow }^{\dagger
},c_{k_{x}1\downarrow },c_{-k_{x}1\downarrow }^{\dagger },c_{k_{x}\bar{1}%
\downarrow },c_{-k_{x}\bar{1}\downarrow }^{\dagger })^{T}$ as
\begin{equation}
\mathcal{H}(k_{x})=\left(
\begin{array}{cc}
A(k_{x}) & B(k_{x}) \\
-B(k_{x}) & A(k_{x})%
\end{array}%
\right) .
\end{equation}%
Here, $A(k_{x})$ and $B(k_{x})$ are $4\times 4$ matrices given by
\begin{equation}
A(k_{x})=\left(
\begin{array}{cccc}
\xi _{k_{x}1} & 0 & 0 & 0 \\
0 & -\xi _{k_{x}1} & 0 & 0 \\
0 & 0 & \xi _{k_{x}\bar{1}} & 0 \\
0 & 0 & 0 & -\xi _{k_{x}\bar{1}}%
\end{array}%
\right) ,
\end{equation}%
\begin{equation}
B(k_{x})=\left(
\begin{array}{cccc}
i2\alpha _{1}\sin k_{x}a_{x} & \Delta _{1} & 0 & \Delta _{c} \\
-\Delta _{1} & -2i\alpha _{1}\sin k_{x}a_{x} & -\Delta _{c} & 0 \\
0 & \Delta _{c} & 2i\alpha _{\bar{1}}\sin k_{x}a_{x} & \Delta _{\bar{1}} \\
-\Delta _{c} & 0 & -\Delta _{\bar{1}} & -2i\alpha _{\bar{1}}\sin k_{x}a_{x}%
\end{array}%
\right) ,
\end{equation}%
In this basis, the time reversal operator is represented as
\begin{equation}
\Theta =\left(
\begin{array}{cc}
0 & I_{4\times 4} \\
-I_{4\times 4} & 0%
\end{array}%
\right) .
\end{equation}%
We can easily confirm that $\Theta ^{\dagger }\mathcal{H}(k_{x})\Theta =\mathcal{H%
}^{\ast }(-k_{x})=\mathcal{H}(k_{x})$. Next, we transform the basis by
unitary matrix $V$  to satisfy
\begin{equation}
V^{\dagger }\Theta V=\left(
\begin{array}{cc}
I_{4\times 4} & 0 \\
0 & -I_{4\times 4}%
\end{array}%
\right),
\end{equation}%
where $V$ can be written as
\begin{equation}
V=\frac{1}{\sqrt{2}}\left(
\begin{array}{cc}
I_{4\times 4} & iI_{4\times 4} \\
iI_{4\times 4} & I_{4\times 4}%
\end{array}%
\right) .
\end{equation}%
As a result, the Hamiltonian is given in the new basis by
\begin{equation}
V^{\dagger }\mathcal{H}(k_{x})V=\left(
\begin{array}{cc}
A(k_{x})+iB(k_{x}) & 0 \\
0 & A(k_{x})-iB(k_{x})%
\end{array}%
\right)\equiv \begin{pmatrix}  \mathcal{H}_{1}(k_{x}) &0 \\
0& \mathcal{H}_{2}(k_{x})
\end{pmatrix}
\end{equation}%
and is decomposed into two independent sectors. In what follows, we focus on only one of two sectors, $\mathcal{H}_{1}(k_{x})$ given by
\begin{equation}
\mathcal{H}_{1}(k_{x})=\left(
\begin{array}{cccc}
\xi _{1}-2\alpha _{1}\sin k_{x}a_{x} & i\Delta _{1} & 0 & i\Delta _{c} \\
-i\Delta _{1} & -\xi _{1}+2\alpha _{1}\sin k_{x}a_{x} & -i\Delta _{c} & 0 \\
0 & i\Delta _{c} & \xi _{\bar{1}}-2\alpha _{\bar{1}}\sin k_{x}a_{x} & i\Delta
_{1} \\
-i\Delta _{c} & 0 & -i\Delta _{1} & -\xi _{\bar{1}}+2\alpha _{\bar{1}}\sin
k_{x}a_{x}%
\end{array}%
\right)
\end{equation}%
The $\mathcal{H}_{1}(k_{x})$ possess the chiral symmetry, $C\mathcal{H}_{1}(k_{x})C=-%
\mathcal{H}_1(k_{x})$, where 
\begin{equation}
C=\left(
\begin{array}{cccc}
0&1&0 & 0 \\
1&0&0&0\\
0&0&0&1\\
0 & 0&1&0%
\end{array}%
\right) .
\end{equation}%
We also define W as
\begin{equation}
W=\left(
\begin{array}{cccc}
1 & 0 & 1 & 0 \\
1 & 0 & -1 & 0 \\
0 & 1 & 0 & 1 \\
0 & 1 & 0 & -1%
\end{array}%
\right)
\end{equation}%
such that
\begin{equation}
W^{\dagger }CW=\left(
\begin{array}{cc}
I_{2\times 2} & 0 \\
0 & -I_{2\times 2}%
\end{array}%
\right) .
\end{equation}%
Using the matrix $W$, $\mathcal{H}_{1}(k_{x})$ is transformed into
\begin{equation}
\mathcal{H}'_{1}(k_{x}) =W^{\dagger }\mathcal{H}_{1}(k_{x})W=\left(
\begin{array}{cc}
0 & h_{1}(k_{x}) \\
h_{1}^{\dagger }(k_{x}) & 0%
\end{array}%
\right) ,
\end{equation}%
where
\begin{equation}
h_{1}(k_{x})=\left(
\begin{array}{cc}
\xi _{k_{x}1}-2\alpha _{1}\sin k_{x}a_{x}-i\Delta _{1} & -i\Delta _{c} \\
-i\Delta _{c} & \xi _{k_{x}\bar{1}}-2\alpha _{\bar{1}}\sin k_{x}a_{x}-i\Delta
_{1}%
\end{array}%
\right)
\end{equation}%
This allows us easily to calculate the determinant $D$ as
\begin{equation}
D_1=\text{det}h_1(k_{x})=(\xi _{k_{x}1}-2\alpha _{1}\sin k_{x}a_{x}-i\Delta
_{1})\times (\xi _{k_{x}\bar{1}}-2\alpha _{\bar{1}}\sin k_{x}a_{x}-i\Delta
_{1})+\Delta _{c}^{2}.
\end{equation}%
The winding number is given by\cite%
{wen}
\begin{equation}
\nu _{1}=\frac{1}{2\pi i}\oint (D_1^{-1}dD_1).
\end{equation}%
The integer $\nu _{1}$ corresponds to the number of the vector composed of the real and imaginary parts of the $D$ wraps around the origin in complex plane when we change the momentum from $k_{x}a_{x}=-\pi$ to  $k_{x}a_{x}=\pi$. We demonstrate that in the topological regime (see  Fig. \ref{fig2}), $\nu _{1}$ is non-zero, confirming the topological protection of the zero energy bound states.

\section{Quasi-classical analysis of the effective model}

\label{bo-san}  In this Appendix we use the effective Hamiltonian given by Eq. (\ref{effective2}) to calculate the ABS spectrum and Josephson current in DQW/DQW junctions. Also, we focus on the most interesting scenario where the left DQW is in the topologically non-trivial phase while the right DQW is in the trivial phase. The system can be viewed as an $s$+$p$/$s$-wave junction.
Since time-reversal symmetry is respected, the Josephson current has the property $I\left(
\varphi \right) =-I\left( -\varphi \right) $. The $s$+$p$/$s$-wave junction
with $\Delta _{s+p}=$ $\left\vert \Delta _{s+p}\right\vert e^{i\varphi }$
and $\Delta _{s}=$ $\left\vert \Delta _{s}\right\vert $ is equivalent to the
$s$/$s$+$p$-wave junction with $\Delta _{s}=\left\vert \Delta
_{s}\right\vert e^{i\varphi }$ and $\Delta _{s+p}=$ $\left\vert \Delta
_{s+p}\right\vert $. In the following calculation, we adopt the latter
convention, such that the Furusaki-Tsukada's formula \cite%
{FT} can be applied directly. We consider a semi-infinite junction wherein an insulating barrier
at $x=0$ separates an $s$-wave superconductor and an $s$+$p$-wave
superconductor. The Hamiltonian of this system is given by%
\begin{align}
&H=H_{s}+H_{I}+H_{s+p}, \\
&H_{s}=\left( k_{x}^{2}/2m-\mu _{0}\right) \tau _{z}-\Delta _{0}e^{i\varphi
}s_{y}\tau _{y}, \\
&H_{I}=H_{\delta }\delta \left( x\right) \tau _{z},\\
&H_{s+p}=\left( k_{x}^{2}/2m-\tilde{\mu}\right) \tau _{z}-\left( \Delta
_{s}+k_{x}s_{y}\Delta _{p}\right) s_{y}\tau _{y},
\end{align}%
where $\varphi $ describes a macroscopic phase difference between two sections
and $\tilde{\mu}=\mu _{s}+2t$. For simplicity, we assume that the chemical
potential is the same in all regions $\mu _{0}=\tilde{\mu}$ and the
interface barrier is modeled by a $\delta $-function with a strength $%
H_{\delta }$. Taking into account the Andreev approximation, we can write
down the eigenmodes of the Hamiltonians $H_{s}$ and $H_{s+p}$. In the $s$-wave
superconductor dominated section  ($x<0$), we find
\begin{subequations}
\begin{alignat}{4}
\psi _{\uparrow ,\pm }^{s,e}\left( x\right) & =\left[ e^{i\varphi
/2},0,0,\gamma _{0}e^{-i\varphi /2}\right] ^{T}e^{\pm ik_{F}x}, & & & & & &
\\
\psi _{\downarrow ,\pm }^{s,e}\left( x\right) & =\left[ 0,e^{i\varphi
/2},-\gamma _{0}e^{-i\varphi /2},0\right] ^{T}e^{\pm ik_{F}x}, & & & & & & \\
\psi _{\uparrow ,\pm }^{s,h}\left( x\right) & =\left[ \gamma _{0}e^{i\varphi
/2},0,0,e^{-i\varphi /2}\right] ^{T}e^{\mp ik_{F}x}, & & & & & & \\
\psi _{\downarrow ,\pm }^{s,h}\left( x\right) & =\left[ 0,\gamma
_{0}e^{i\varphi /2},-e^{-i\varphi /2},0\right] ^{T}e^{\mp ik_{F}x}, & & & &
& &
\end{alignat}%
where we have defined $k_{F}=\sqrt{2m\tilde{\mu}}$ and $\gamma _{0}=\Delta
_{0}(E+\sqrt{E^{2}-\Delta _{0}^{2}})^{-1}$.  The quasiparticle energy $E$ is
measured from the chemical potential and the subscript \textquotedblleft $+$%
" (\textquotedblleft $-$") stands for the right-going (left-going)
solutions. For the $s$+$p$-wave superconductor dominated section ($x>0$), we only consider the
right-going solutions of $H_{s+p}$ given by
\end{subequations}
\begin{subequations}
\begin{alignat}{4}
\psi _{1}^{s+p,e}\left( x\right) & =\left[ 1,-i,i\gamma _{1},\gamma _{1}%
\right] ^{T}e^{ik_{F}x}, & & & & & & \\
\psi _{2}^{s+p,e}\left( x\right) & =\left[ 1,i,-i\gamma _{2},\gamma _{2}%
\right] ^{T}e^{ik_{F}x}, & & & & & & \\
\psi _{1}^{s+p,h}\left( x\right) & =\left[ \gamma _{1},i\gamma _{1},-i,1%
\right] ^{T}e^{-ik_{F}x}, & & & & & & \\
\psi _{2}^{s+p,h}\left( x\right) & =\left[ \gamma _{2},-i\gamma _{2},i,1%
\right] ^{T}e^{-ik_{F}x}, & & & & & &
\end{alignat}%
with $\gamma _{1\left( 2\right) }=\left( \Delta _{s}\mp k_{F}\Delta
_{p}\right) [E+\sqrt{E^{2}-\left( \Delta _{s}\mp k_{F}\Delta _{p}\right) ^{2}%
}]^{-1}$ reflecting the existence of two superconducting gaps $\Delta
_{1}=\left\vert \Delta _{s}-k_{F}\Delta _{p}\right\vert $ and $\Delta
_{2}=\left\vert \Delta _{s}+k_{F}\Delta _{p}\right\vert $. In order to
obtain the ABSs, we consider the following wave function
\end{subequations}
\begin{align}
&\Psi _{s}\left( x\right) =\sum_{\sigma =\uparrow ,\downarrow }r_{\sigma
}^{s,e}\psi _{\sigma ,-}^{s,e}\left( x\right) +r_{\sigma }^{s,h}\psi
_{\sigma ,-}^{s,h}\left( x\right) , \ \ \ x<0\\
&\Psi _{s+p}\left( x\right) =\sum_{\nu =1,2}r_{\nu }^{s+p,e}\psi _{\nu
}^{s+p,e}\left( x\right) +r_{\nu }^{s+p,h}\psi _{\nu }^{s+p,h}\left(
x\right),  \ \ \ x>0.
\end{align}
The scattering coefficients $r_{\sigma }^{s,e}$, $%
r_{\sigma }^{s,h}$, $r_{\nu }^{s+p,e}$ and $r_{\nu }^{s+p,h}$ are chosen such that
the boundary condition at $x=0$ are satisfied,
\begin{align}
&\Psi _{s}\left( 0\right) =\Psi _{s+p}\left( 0\right) , \\
&\partial _{x}\Psi
_{s+p}\left( 0\right) -\partial _{x}\Psi _{s}\left( 0\right) =2mH_{\delta
}\Psi _{s}\left( 0\right) .  \label{LBbound}
\end{align}%
The discrete ABSs can be determined by the condition $\det M=0$, where the matrix $M$ is defined as
\begin{align}
&M=\left[
\begin{array}{cc}
M_{1} & M_{2} \\
M_{3} & M_{4}%
\end{array}%
\right] , \\
&M_{1}=\left[
\begin{array}{cccc}
0 & \gamma _{0}e^{i\varphi /2} & 0 & e^{i\varphi /2} \\
\gamma _{0}e^{i\varphi /2} & 0 & e^{i\varphi /2} & 0 \\
-e^{-i\varphi /2} & 0 & -\gamma _{0}e^{-i\varphi /2} & 0 \\
0 & e^{-i\varphi /2} & 0 & \gamma _{0}e^{-i\varphi /2}%
\end{array}%
\right] ,\\
&M_{2}=\left[
\begin{array}{cccc}
-1 & -\gamma _{1} & -1 & -\gamma _{2} \\
i & -i\gamma _{1} & -i & i\gamma _{2} \\
-i\gamma _{1} & i & i\gamma _{2} & -i \\
-\gamma _{1} & -1 & -\gamma _{2} & -1%
\end{array}%
\right] ,\\
&M_{3}=\left[
\begin{array}{cccc}
0 & (1-iZ)\gamma _{0}e^{i\varphi /2} & 0 & (-1-iZ)e^{i\varphi /2} \\
(1-iZ)\gamma _{0}e^{i\varphi /2} & 0 & (-1-iZ)e^{i\varphi /2} & 0 \\
-(1-iZ)e^{-i\varphi /2} & 0 & (1+iZ)\gamma _{0}e^{-i\varphi /2} & 0 \\
0 & (1-iZ)e^{-i\varphi /2} & 0 & (-1-iZ)\gamma _{0}e^{-i\varphi /2}%
\end{array}%
\right] ,\\
&M_{4}=\left[
\begin{array}{cccc}
-1 & \gamma _{1} & -1 & \gamma _{2} \\
i & i\gamma _{1} & -i & -i\gamma _{2} \\
-i\gamma _{1} & -i & i\gamma _{2} & i \\
-\gamma _{1} & 1 & -\gamma _{2} & 1%
\end{array}%
\right],
\end{align}%
where we used the notation $Z=2mH_{\delta }/k_{F}$.

At the next step, we calculate the Josephson current using Furusaki-Tsukada's formula \cite{FT} by considering the incoming electrons from the left side,
\begin{subequations}
\begin{alignat}{4}
\Psi _{\uparrow }^{s,e}\left( x\right) & =\psi _{\uparrow ,+}^{s,e}\left(
x\right) +\sum\nolimits_{\sigma }\left[ r_{\uparrow \sigma }^{ee}\psi
_{\sigma ,-}^{s,e}\left( x\right) +r_{\uparrow \sigma }^{eh}\psi _{\sigma
,-}^{s,h}\left( x\right) \right] , & & & & & & \\
\Psi _{\downarrow }^{s,e}\left( x\right) & =\psi _{\downarrow
,+}^{s,e}\left( x\right) +\sum\nolimits_{\sigma }\left[ r_{\downarrow \sigma
}^{ee}\psi _{\sigma ,-}^{s,e}\left( x\right) +r_{\downarrow \sigma
}^{eh}\psi _{\sigma ,-}^{s,h}\left( x\right) \right] , & & & & & & \\
\Psi _{\uparrow }^{s,h}\left( x\right) & =\psi _{\uparrow ,+}^{s,h}\left(
x\right) +\sum\nolimits_{\sigma }\left[ r_{\uparrow \sigma }^{he}\psi
_{\sigma ,-}^{s,e}\left( x\right) +r_{\uparrow \sigma }^{hh}\psi _{\sigma
,-}^{s,h}\left( x\right) \right] , & & & & & & \\
\Psi _{\downarrow }^{s,h}\left( x\right) & =\psi _{\downarrow
,+}^{s,h}\left( x\right) +\sum\nolimits_{\sigma }\left[ r_{\downarrow \sigma
}^{he}\psi _{\sigma ,-}^{s,e}\left( x\right) +r_{\downarrow \sigma
}^{hh}\psi _{\sigma ,-}^{s,h}\left( x\right) \right] . & & & & & &
\end{alignat}%
All the coefficients can be found from the same boundary conditions given by Eq. (\ref%
{LBbound}). The Josephson current is given by \cite{FT}
\end{subequations}
\begin{equation}
I=\frac{e\Delta _{0}}{2}k_{B}T\sum_{\omega _{n},\sigma }\frac{\mathrm{sgn}%
(\omega _{n})}{\Omega _{n}}\left[ r_{\sigma \sigma }^{eh}\left( i\omega
_{n}\right) -r_{\sigma \sigma }^{he}\left( i\omega _{n}\right) \right] ,
\end{equation}%
where $r_{\sigma \sigma }^{eh}\left( i\omega _{n}\right) $ and $r_{\sigma
\sigma }^{he}\left( i\omega _{n}\right) $ are obtained by the analytical
continuation $E\rightarrow i\omega _{n}$ of $r_{\sigma \sigma }^{eh}\left(
E\right) $ and $r_{\sigma \sigma }^{he}\left( E\right) $.
The Matsubara frequency $\omega _{n}$ is defined as $\omega _{n}=\pi k_{B}T\left( 2n+1\right) $ for $%
n=0,\pm 1,\pm 2\cdot \cdot \cdot $, and $\Omega _{n}=\sqrt{\omega
_{n}^{2}+\Delta _{0}^{2}}$. Here, we work in the low temperature limit and neglect the temperature dependence of the superconducting gap.

 \end{widetext}

\bibliography{nanowire1}
\end{document}